\documentstyle[prb,aps,preprint]{revtex}
\begin{document}
\title{Ground states of a one-dimensional lattice gas model with
an infinite range nonconvex interaction. A numerical study}
\author{Cz. Oleksy and J. Lorenc}
\address{Institute of Theoretical Physics, University of Wroc{\l}aw
Pl. Maksa Borna 9, 50-204 Wroc{\l}aw, Poland}
\date{\today}
\maketitle
\draft
\begin{abstract}
We consider a lattice gas model with an infinite pairwise nonconvex
total interaction of the form
\[ 
V(r)=\frac{J}{r^2} + A\frac{\cos (2k_Far+\phi )}{r}\;.
\]
This one-dimensional interaction might account, for example, for
adsorption of alkaline elements on W(112) and Mo(112). The first
term describes the effective dipole-dipole interaction while the other
one the indirect (oscillatory) interaction; $J$, $A$, and $\phi$
are the model parameters, whereas $k_F$ stands for the wavevector of
electrons at the Fermi surface and $a$ is a lattice constant.
We search for the (periodic) ground states. To solve this difficult
problem we have applied a novel numerical method to accelerate 
the convergence of Fourier series.
A competition between the dipole-dipole and indirect interactions 
turns out to be very important. We have found that the reduced
chemical potential $\mu/J$ versus $A/J$ phase diagrams contain a
region $0.1 \leq A/J \leq 1.5$  dominated by several phases only with
periods up to nine lattice constants. Of course, the resulting
sequence of phases (for fixed $A/J$)   depends on the wavevector
$k_F$ and the phase shift $\phi$.
The remaining phase diagram reveals a complex structure of
usually long periodic phases.
We conjecture, based on the above results, that 
qua\-si\--one-\-di\-men\-sio\-nal surface states might be responsible 
for experimentally observed ordered phases at the (112) surface of 
tungsten and  molybdenum.
\end{abstract}

\section{Introduction}

There has been a widespread interest in structures and phase
transitions in metal submonolayers chemisorbed on metal surfaces
(e.g., for a review see Refs.\ \onlinecite{ref1,ref2}). In particular,
many alkaline, alkaline-earth, and rare-earth elements adsorbed on
(112) faces of tungsten and
molybdenum form, for low coverages, ordered structures consisting of
linear chains of adatoms. Recently, a lattice gas model has been
proposed to account for these structures.\cite{ref3}
The results indicate that a formation of the linear chain submonolayer
structures might be due to
the competing (repulsive) dipole-dipole interaction and the long-range
(oscillatory) indirect interaction between adsorbates. The indirect
interaction was assumed to be mediated by the quasi-one-dimensional 
valence electronic states of the underlaying substrate and was closely 
related to the existence of nearly flattened segments of the Fermi 
surface of W or Mo.\cite{ref2,ref3}
It has been demonstrated 
explicitly that the resulting structures are very sensitive to a 
competition between  the dipole-dipole and indirect
interactions.\cite{ref3}

It is the intention of this paper to understand the role of the
interactions in determining the ordered structures. Since
the general problem turns out to be quite complex, it is useful to
focus on a simple model which contains the essential ingredients.
We have considered an effective one-dimensional lattice gas model
with the competing infinite range (convex) dipole-dipole and
(nonconvex) indirect interactions and we study its ground-state phase
diagrams. Our analysis is based on a novel numerical method to
accelerate the convergence of Fourier series \cite{ref4} in order to
take into account the interactions of the infinite range.
In Sec.\ \ref{s2} we introduce a lattice gas model. The periodic
ground states have been calculated numerically and  the resulting
phase diagrams are discussed in Sec.\ \ref{s3}. A possible role of
surface states is considered in Sec.\ \ref{s4}. The obtained results
are summarized together with the concluding remarks in Sec.\ \ref{s5}.

\section{The model}\label{s2}
It is well known that the lattice gas models are quite useful in
studying overlayer structures and their properties (e.g.,
Refs.\ \onlinecite{ref5,ref6}). Here, we consider an effective
one-dimensional lattice gas model with the following grand canonical
ensemble Hamiltonian
\begin{equation}\label{hamiltonian}
{H}=\sum_{i}\sum_{r=1}^\infty
 V(r)n_{i}n_{i+r}-\mu\sum_{i}n_{i} \;,
\end{equation}
where $n_{i}=1$ or $0$ depending on whether the {\em ith} site
of the one-dimensional lattice is occupied or not by an adatom; 
$V(r)$  is an interaction between a pair of adatoms separated by
distance $r$ and $\mu$ is the chemical potential which controls
the coverage of adatoms. The distance $r$ is measured in units of a
lattice constant.

We assume the pairwise total interaction $V(r)$ in the form
\begin{equation}\label{potential}
V(r)=\frac{J}{r^2} + A\frac{\cos (2k_Far+\phi )}{r}\;.
\end{equation}
The first term in Eq.\ (\ref{potential})
describes the effective di\-po\-le-di\-po\-le interaction in one
dimension
relevant to linear chain structures while the other one represents the
indirect
(oscillatory) interaction (see also Refs.\ \onlinecite{ref2,ref3});
$J$, $A$, and $\phi$  are the model parameters, whereas $k_F$
stands 
for the wavevector of electrons at the Fermi surface and $a$ is a 
lattice constant of the lattice gas model.

In the following, we restrict ourselves to periodic configurations of
$n$'s, i.e., there is a period $p$ (positive integer) such that
$n_{j+p} = n_{j}$ for any integer $j$. These are called $p$-periodic
configurations of $n$'s.

Now, the energy per site of a given $p$-periodic configuration
$n_1, \ldots, n_p$ can be written in the form

\begin{equation}\label{ener}
 E[n_1, \ldots, n_p] = \sum_{r=1}^{p}E_{p}(r)\sum_{i=1}^{p} n_i
n_{i+r} - \mu \theta \;,
\end{equation}

where

\begin{equation}\label{ener1}
E_p(r)= J D_p(r) + A C_p(r) \;,
\end{equation}

\begin{equation}\label{ener2}
D_p(r)=\frac{1}{p^3}\sum_{s=0}^{\infty}\frac{1}{(s + r/p)^2} \;,
\end{equation}

\begin{equation}\label{ener3}
C_p(r)=\frac{1}{p^2}\sum_{s=0}^{\infty}\frac{\cos [x_p(s + r/p)+
\phi ]}{s + r/p} \;,
\end{equation}

\begin{equation}\label{ener4}
x_p=2 k_F a p \;,
\end{equation}

\begin{equation}\label{ener5}
\theta=\frac{1}{p}\sum_{i=1}^{p}n_i \;.
\end{equation}

Let us note that the infinite range of the total interaction $V(r)$ 
requires the exact summations in Eqs.\ (\ref{ener2}) and
(\ref{ener3}). 
This numerical problem  will be discussed in the next section.

\section{The ground states}\label{s3}

It is difficult to determine the ground states of the model
Hamiltonian $H$ rigorously because the total interparticle
interaction
$V(r)$ is {\em nonconvex} and we assume it to have the infinite
range.
This assumption seems to be necessary for the indirect (oscillatory)
interaction is truly long-ranged. For a class of infinite-range
interactions which are convex, positive, etc., it has been possible
to find, in a one-dimensional case, all the ground states for any
rational $0 < q/p < 1$, where $q$ and $p$ are integers with no common
multipliers.\cite{ref7,ref8,ref9,ref10}
The effects of nonconvexity on the ground
states might be discussed, in principle, within a new general method
of Griffiths and Chou.\cite{ref11,ref12} The method makes it possible
to find
the ground state configuration and the energy of one-dimensional
systems
by studying the corresponding nonlinear equation whose solution is an
effective potential. Unfortunately, the method of Griffths and Chou
does not seem to work
in the present case because of the infinite range of the slowly
decaying total interaction [$V(r) \sim r^{-1}$ for large $r$] which
makes it practically
impossible to solve the eigenvalue equation. Note, however, that the
method works well for interactions decaying
exponentially.\cite{ref13}

\subsection{A numerical search for the ground states}

In order to find the ground states we consider, instead of all
possible  configurations, only  $p$-periodic configurations
with the energy per site given by Eq.\ (\ref{ener}). 
Recently, we have proposed a numerical procedure
to find such ground states.\cite{ref3} It consists of generating
numerically all $p$-periodic configurations of $n$'s with 
$p=1, \ldots, p_{max}$ by using
the bit representation of integers from the interval $[2^{p-1}, 2^p]$.
The number of configurations could then be reduced by making use of
the  particle-hole symmetry \cite{ref5} as well as the translational
and/or inversion symmetries. In this way, for example, the total
number of $p$-periodic
configurations for $p_{max}=23$, $2^{23}$, has been reduced to
$181884$  distinct configurations, i.e., by a factor of 46. Next,
we check explicitly which configurations of $n$'s
afford the minimal value to the corresponding $E[n_1, \ldots, n_p]$ 
$(p=1, \ldots, p_{max})$, Eq.\ (\ref{ener}). However, this requires  
calculations of $D_p(r)$ and $C_p(r)$, Eq.\ (\ref{ener2}) and 
Eq.\ (\ref{ener3}), for $r=1, \ldots, p$ and $p=1, \ldots, p_{max}$. 
Here, we note that the infinite series result 
from the infinite range of the interparticle total interaction $V(r)$. 
It is not possible, to our knowledge, to calculate most of the sums of 
the corresponding infinite series analytically. Even a numerical 
calculation presents a problem due to a quite slow convergence of the
series. 

We have calculated numerically
$D_p(r)$, Eq.\ (\ref{ener2}), by using an analytic method of the
convergence acceleration of the series.\cite{ref14} It is interesting
to observe that
for the $p$-periodic configurations we obtain the same contribution to 
$E[n_1, \ldots, n_p]$, Eq.\ (\ref{ener}), by replacing $D_p(r)$ via 
$[D_p(r) + D_p(p-r)]/2$ (see Ref.\ \onlinecite{ref15}). The numerical
calculations of $C_p(r)$, Eq.\ (\ref{ener3}) and Eq.\ (\ref{ener4}),
are more difficult for one can check explicitly that the standard
methods to accelerate the convergence of the series do not work
\cite{ref16}
(especially for $mod_{2\pi}x_p$ close to $0$ or $2\pi$).
To overcome this problem we have applied a novel method \cite{ref4}
which introduces the so-called initial transformation and is followed
by the $\epsilon$ algorithm for the associated complex Fourier series.
The results are numerically exact (in double precision).

\subsection{The ground-state phase diagrams}

The ground-state configuration and the corresponding energy per site 
depend on the chemical potential, $\mu$, and the model parameters,
such as, $J$, $A$, $\phi$, $k_F$, $a$, and $p_{max}$. In principle,
calculations should be performed for $p_{max}$ going to infinity to
allow for any periodic structure. In practice, we have carried out
calculations
for $p_{max}=23$ as a compromise between the computing time and more 
refined results. Our numerical tests show that to understand the role
of the nonconvex interparticle total interaction $V(r)$ in
determining the ground states, no additional significant insight can
be achieved by extending the numerical computations to higher values
of $p_{max}$.

The best way to discuss a competition between the effective
dipole-dipole
and indirect interactions (or, equivalently, the role of nonconvexity
) is to consider ground-state phase diagrams: the reduced chemical
potential  $\mu/J$ versus model parameters $A/J, \phi, k_F$.
In the following calculations we shall assume $a=3.16 \sqrt 3 /2\;\AA$.

In constructing the phase diagrams we made use of the particle-hole
symmetry, i.e., we present only the $p$-per\-io\-dic ground states
with $\theta\le 1/2$ for $\mu < \mu_c~=~\sum_{r=1}^{\infty}V(r)$.
Moreover, it is sufficient to consider only $k_F\in(0, \pi/a]$ because
$E[n_1,\ldots, n_p]$, Eq.\ (\ref{ener}), is periodic with respect to
$k_F$  (with a period of $\pi/a$) and it is invariant under the
transformation:
($k_F,\phi$) onto ($k_F',\phi'$) = ($\pi/a - k_F, -\phi$).
We also assume $\phi\in(-\pi,\pi]$.

Most of the $\mu$ versus $x$ phase diagrams ($x$ = $A/J$, $\phi$,
$k_F$)
turned out to have rather complex topology (up to 100 lines seperating 
distinct phases). In order to determine precisely the regions of
stability  of phases we find sequences of the ground states at $x$ and
$x+\Delta x$, respectively.
Usually, we assume $\Delta x$ to be $1/50$ of the whole range of $x$. 
Then, if the seqences are different, we make use of the bisection
method,  otherwise we continue our scan in $x$.

Now, we shall discuss the results with a special emphasis on the role
the indirect (nonconvex) interaction plays in determining the ground
states. As the reference results we consider the  ground states for
$A=0$  which form,
on the ($\theta,\mu$) phase diagram, the complete devil's staircase
with  a rather involved fractal behavior.\cite{ref7,ref8,ref9,ref10}

We use standard notation for the ground state, i.e., $q/p$, which
means $q$ occupied sites in a unit cell of $p$ sites. At the same time
$\theta=q/p$  and, in the present study, we are not interested in what
are the actual structures of the ground states.
As we shall see, the calculated phase diagrams turned out to be quite
complex. Therefore, in order to make them more clear we do not attach
$q/p$ labels to  small regions of the phase diagrams usually
corresponding to the ground states with lager periodicities.

\subsubsection{Dependence on $\phi$}

We start with the nonconvex indirect interaction only ($J=0$). A
typical chemical potential $\mu$ versus phase $\phi/\pi$
ground-state phase diagram is shown in Fig.\ \ref{fig1}(a). The value
$k_F=0.82\;\AA^{-1}$ was chosen in
view of the subsequent discussion. The broken line depicts the
particle-hole symmetry line $\mu_c=\sum_{r=1}^{\infty}V(r)$.
%\begin{figure}[]
%\centering
%\mbox{
%\epsfxsize=6cm
%\epsfbox[160 150 450 700]{fig1.ps}
%}
%\caption{
%The ground-state phase diagrams. The broken line depicts the
%particle-hole
%symmetry line $\mu_c=\sum_{r=1}^{\infty}V(r)$. (a) $\mu$ vs
%$\phi/\pi$ for $J=0$, $A=1$, $k_F=0.82\;\AA^{-1}$.
%(b) $\mu/J$ vs $\phi/\pi$ for  $A/J=0.5$, $k_F=0.82\;\AA^{-1}$.
%}\label{fig1}
%\end{figure}
Here, $0/1$ and $1/1$ denote the low-density
($\theta=0$) and high-density ($\theta=1$) disordered ground-states, 
respectively. 

Our results demonstrate clearly that the ground-state phase
diagram is very sensitive to $p_{max}$ for $0.5 < |\phi/\pi| < 1$.
The part of the phase diagram is not ``stable'' against formation of 
the longer-periodic ground states for larger values of $p_{max}$. 
This particular behavior might be attributed to the fact that 
the function $C_p(r)$, Eq.\ (\ref{ener3}) and Eq.\ (\ref{ener4}), 
has the smaller value the closer
$mod_{2\pi}x_p$ is to $0$ or $2\pi$. Thus, for a fixed $p_{max}$ one
can always find such $p>p_{max}$ for which the above condition is
satisfied and
this could be easily tested numerically. This is the reason for not 
discussing the part of phase diagrams.

The situation, however, is quite different for $|\phi/\pi|<0.5$.
Figure\ \ref{fig1}(a) shows that apart from the $1/2$ and $1/3$
ordered ground
states there is a coexistence of the low-density ($\theta=0$) and
high-density
($\theta=1$) disordered ground states. It is due to an effective 
attraction of the nonconvex indirect interaction. Inclusion of the
repulsive convex dipole-dipole interaction changes this part of the 
phase diagram and some of the typical results are presented in
Fig.\ \ref{fig1}(b). 
The dependence on $\phi$ is very important from a physical point of
view
because by changing its value one can induce distinct sequences of 
the ground states. It is worth to notice that relatively large parts
of  the phase diagram, Fig.\ \ref{fig1}(b), are occupied by the 
ground states
having low periodicities $p$, like $2, 3, 4, 5,$ or $6$.

\subsubsection{Dependence on $A/J$}

The ground states depend on the strength of the nonconvex interaction
$A$ in a crucial way. This is demonstrated explicitly in
Fig.\ \ref{fig2}(a) and Fig.\ \ref{fig2}(b).
%\begin{figure}[]
%\centering
%\mbox{
%\epsfxsize=6cm
%\epsfbox[160 150 450 700]{fig2.ps}
%}
%\caption{
%The ground-state phase diagrams $\mu/J$ vs $A/J$ for
%$k_F=0.82\;\AA^{-1}$. (a) $\phi=-0.3\pi$. (b) $\phi=0$.
%}\label{fig2}
%\end{figure}
It is easy to see that the devil's sequence of the ground states
($A=0$) is destroyed even for a small value of $A/J$, leaving only a
number of lower periodic phases. More precisely, for $0.1<A/J<1.5$,
the phase diagram is dominated by such phases. The numerical tests
show that this topology is not affected by $p_{max}$. We believe that
this is a ``physical'' region which might account for the ground-state
structures and/or phase transitions in linear chain 
structures,\cite{ref1,ref2,ref3} at least. Indeed, following our 
previous model of adsorption of lithium on W(112) and Mo(112)
(see Ref.\ \onlinecite{ref3}), we have estimated $A/J$ for
the two cases. We find $A/J\approx 0.3$ ($\phi\approx 0.3\pi$) for
Li/W(112) with
$k_F=0.41\;\AA^{-1}$ and $A/J\approx1.3$ ($\phi\approx -0.1\pi$) for
Li/Mo(112) with $k_F=0.47\;\AA^{-1}$. The corresponding sequences of
the ground states have been
denoted by the vertical dashed-dotted lines in Fig.\ \ref{fig3}(a) and
Fig.\ \ref{fig3}(b),  respectively. A very good agreement with 
previous results, Ref.\ \onlinecite{ref3},
is in favour of the present one-dimensional model as far as the linear
chain  structures are
concerned. Let us observe that for sufficiently large $A/J$ the lower
periodic phases are replaced by a number of the narrow ground states 
having large periodicities.
%\begin{figure}[]
%\centering
%\mbox{
%\epsfxsize=6cm
%\epsfbox[160 150 450 700]{fig3.ps}
%}
%\caption{
%The ground-state phase diagrams $\mu/J$ vs $A/J$. (a)
%$\phi=0.3\pi$, $k_F=0.41\;\AA^{-1}$. The vertical dashed-dotted line
%at $A/J=0.3$ corresponds to the case of Li/W(112).
%(b)
%$\phi=-0.1\pi$, $k_F=0.47\;\AA^{-1}$. The vertical dashed-dotted line
%at $A/J=1.3$ corresponds to the case of Li/Mo(112).
%}\label{fig3}
%\end{figure}

\subsubsection{Dependence on $k_F$}

This dependence is important because of the underlying physics leading
to the indirect interaction (virtual bulk and/or surface electronic
states).\cite{ref1,ref2}
However, to get more insight into properties of the
ground states one can consider $k_F$ as a parameter and some of the 
results obtained for such a case are shown in 
Fig.\ \ref{fig4} and Fig.\ \ref{fig5}.
%\begin{figure}[]
%\centering
%\mbox{
%\epsfxsize=6cm
%\epsfbox[160 270 450 560]{fig4.ps}
%}
%\caption{
%The ground-state phase diagram $\mu$ vs $k_F$ for $J=0$, $A=1$, and
%$\phi=0$.
%}\label{fig4}
%\end{figure}

Note that the upper scale, $\pi/a-k_F$,
is related to the lower one, $k_F$, for $\phi=0$. And, as
before, the broken line denotes a particle-hole symmetry line. The 
ground-state phase diagram for the nonconvex indirect interaction
only is shown in Fig.\ \ref{fig4}.
%\begin{figure}[]
%\centering
%\mbox{
%\epsfxsize=6cm
%\epsfbox[160 140 450 700]{fig5.ps}
%}
%\caption{
%The ground-state phase diagrams $\mu/J$ vs $k_F$ for $A/J =0.2$. 
%(a) $\phi=0$. (b) $\phi=-0.3\pi$.
%}\label{fig5}
%\end{figure}
Let us note that for $k_F \rightarrow  0$ one restores
the devil's staircase because the interaction $V(r)$ becomes convex.
Modifications to the phase diagram in Fig.\ \ref{fig4} due to a 
competition between the interactions are presented 
in Fig.\ \ref{fig5}(a) and Fig.\ \ref{fig5}(b).
Again, the presented phase diagrams show relatively large, stable
regions of the ground states with lower periodicities. It can be seen
that the role
of the phase, $\phi$, is to ``shift'' the corresponding phase diagram
as shown in Fig.\ \ref{fig5}(a) and Fig.\ \ref{fig5}(b).

\section{A possible role of surface states}\label{s4}

The obtained results suggest that a given sequence of the ground
states could be described by different $k_F$ wavevectors with
appropriately
chosen amplitudes $A$ and phases $\phi$. For example, one of
the possibilities may be connected with the transformation
($k_F',\phi'$) = ($\pi/a - k_F,-\phi$) for a fixed value of $A$
(see also Sec.\ \ref{s3}).
This observation leads us to a conjecture concerning a possible role
of  {\em surface states} in understanding a formation of the linear
chain structures. Indeed, the indirect interaction between adatoms via
virtual electrons from  quasi-one-dimensional surface states has the
same form, as we have already discussed
(cf. Refs.\ \onlinecite{ref2,ref17,ref18,ref19}).
Now, the wavevector $k_F$ is determined from the Fermi lines 
corresponding to partially filled surface-state band(s) and it can
be different from the one relevant to the bulk electronic states.

To our best knowledge, there are neither experimental nor theoretical
results concerning partially filled surface-state band(s) of the (112)
surface of W and Mo. However, a recent suggestion by Tosatti
\cite{ref20} as well as interpretation of the LEED data obtained for
adsorption of lithium on W(112), Mo(112), and Ta(112) \cite{ref21}
lead to a conclusion that surface states or surface resonances could
exist on W(112) and Mo(112).
Of course, this conclusion has  to be confirmed by the corresponding
experiments and selfconsistent calculations following, for
example, Ref.\ \onlinecite{ref22}.

In general, the adsorption can substantially influence  surface states
by changing their band(s) structure and/or the surface Brillouin zone.
Consequently, the form of the indirect interaction mediated by
surface states \cite{ref2} could be changed. In our case, however, we
study the model relevant to the low coverage linear chain structures
and it is plausible, therefore, that the influence on surface states
can be neglected.

Our results seem to be consistent with the conjecture concerning
surface states. For example, in order to explain a Sr/Mo(112)
adsorption system at low coverages, one has to consider
$k_F \approx 0.82\;\AA^{-1}$ and not the wavevector
$k_F=0.47\;\AA^{-1}$
connected with the bulk electronic states. Also a Li/Mo(112) system
could equally well be described by using $k_F \approx 0.82\;\AA^{-1}$
[see Fig.\ \ref{fig2}(a)].

\section{Discussion and conclusions}\label{s5}

The paper analyses the ground states of a one-di\-men\-sio\-nal
lattice gas model with the infinite range pairwise interaction being
the result of a
competition between the convex dipole-dipole and nonconvex indirect 
interactions. The ground-state  phase diagrams depend on three model 
parameters: $A/J$, $k_F$, and $\phi$. Their complicated, at first
sight, structures have, however, quite simple explanation. Indeed,
\begin{enumerate}
\item By increasing $A/J$ from $0.1$ to $1.5$ (approximately) one 
{\em reduces} sequences of the ground states to those which contain
only several lower-periodic phases. These phases exist in relatively
large, stable regions, whereas the longer periodic phases enter
only in narrow ranges in between (Figs.\ \ref{fig2} and \ref{fig3}).
\item The wavevector, $k_F$, changes sequences of the ground states.
\item The change in the phase, $\phi$, results effectively in a
``shift'' of the corresponding phase diagram in $k_F$ 
[Figs.\ \ref{fig5}(a) and \ref{fig5}(b)].
\end{enumerate}

It seems that the present model can describe qualitatively linear
chain phases adsorbed at low temperatures on the (112) face of W or
Mo.\cite{ref3}
The model parameters have a definite physical meaning and
distinct  sequences of phases (the ground states) are governed by
different values of $A/J$ and $\phi$ which, in turn, depend on a
particular adsorbate-substrate system. Of course, the wavevector $k_F$
is closely related to a mechanism of the indirect interaction (bulk or
surface electronic states).

Different linear chain structures, observed experimentally at $77\;K$ 
on the (112) surface,\cite{ref2}  might be described within our model
by assuming one value of $k_F$ with different values of $\phi$.
The remaining parameter, $A/J$, is to be taken from the range (0.1,
1.5). For Li/W(112)
and Li/Mo(112) we can estimate $A/J$ (note that $\phi$ and $k_F$
are chosen as in Ref.\ \onlinecite{ref3}) and the results are shown
in Fig.\ \ref{fig3}(a) and Fig.\ \ref{fig3}(b), respectively. 
A qualitative agreement is
in  favour of the present simplified model. It is interesting to
observe that for
$k_F=0.82\;\AA^{-1}$ (or $\pi/a-0.82\;\AA^{-1}$), Fig.\ \ref{fig5}(a),
we find the ground
states with periods 2, 5, and 9 (in units of a lattice constant, $a$). 
The higher periodic ground states enter only in small
intervals [see also Fig.\ \ref{fig2}(b)]. Our numerical study
reveals that only small 
deviations of the model parameters, i.e., $A/J$, $k_F$, and $\phi$,
are 
possible in order to simulate the above sequence of the ground states.
The 
case might be related to strontium submonolayer linear chain
structures on Mo(112) \cite{ref2} observed experimentally at
$T=77\;K$. Moreover, the physical mechanism behind the form of the
indirect interaction does not
seem to be connected with the virtual bulk electronic states
\cite{ref23,ref24} because this would require $k_F=0.47\;\AA^{-1}$ .
Instead, we conjecture that (one-dimensional) {\em surface} electronic
states might
be responsible for the ground states in the case of Sr/Mo(112). The
wavevector $k_F=0.82\;\AA^{-1}$ (or $\pi/a-0.82\;\AA^{-1}$) would be
determined by a band of surface states at the Fermi surface.
(For details concerning the indirect interaction induced
by surface states see, for example, Ref.\ \onlinecite{ref2}).
We would like to stress that the adsorption of lithium on W(112) 
could equally well be explained by the existence of surface
states with the corresponding wavevector, $k_F'$ and $\phi'$
related  by the transformation $(k_F',\phi') = (\pi/a - k_F,
-\phi)$.

It would be interesting to verify the existence of partially filled
surface
state band(s) on W(112) and Mo(112). Moreover, a knowledge of the 
corresponding $k_F$ could help to construct a more realistic 
two-dimen\-sional model of phase transitions.

We believe that our results for the ground states can be treated as a 
starting point
towards understanding experimental data at $77\;K$. A more advanced
theory
(Monte Carlo simulation, real-space or finite-size renormalization
group,
cf. Refs.\ \onlinecite{ref25,ref26}) is needed to understand the
($T,\theta$) phase
diagram and to compare the underlaying experimental data quantatively.
This, however, would also require extension of the present model to
two  dimensions to make it more realistic.

\acknowledgments

It is a pleasure to thank Professor Stefan Paszkowski for discussions
on methods for accelerating the convergence of series. One of us
(Cz. O.)
is indebted to Professor Erio Tosatti for discussions regarding
surface
states. We are also grateful to Professor Jan Ko{\l}aczkiewicz for his
interest in the present work.

\begin{figure}[h]
\caption{
The ground-state phase diagrams. The broken line depicts the
particle-hole
symmetry line $\mu_c=\sum_{r=1}^{\infty}V(r)$. (a) $\mu$ vs
$\phi/\pi$ for $J=0$, $A=1$, $k_F=0.82\;\AA^{-1}$.
(b) $\mu/J$ vs $\phi/\pi$ for  $A/J=0.5$, $k_F=0.82\;\AA^{-1}$.
}\label{fig1}
\end{figure}
\begin{figure}[h]
\caption{
The ground-state phase diagrams $\mu/J$ vs $A/J$ for
$k_F=0.82\;\AA^{-1}$. (a) $\phi=-0.3\pi$. (b) $\phi=0$.
}\label{fig2}
\end{figure}
\begin{figure}[h]
\caption{
The ground-state phase diagrams $\mu/J$ vs $A/J$. (a)
$\phi=0.3\pi$, $k_F=0.41\;\AA^{-1}$. The vertical dashed-dotted line
at $A/J=0.3$ corresponds to the case of Li/W(112).
(b)
$\phi=-0.1\pi$, $k_F=0.47\;\AA^{-1}$. The vertical dashed-dotted line
at $A/J=1.3$ corresponds to the case of Li/Mo(112).
}\label{fig3}
\end{figure}
\begin{figure}[h]
\caption{
The ground-state phase diagram $\mu$ vs $k_F$ for $J=0$, $A=1$, and
$\phi=0$.
}\label{fig4}
\end{figure}
\begin{figure}[h]
\caption{
The ground-state phase diagrams $\mu/J$ vs $k_F$ for $A/J =0.2$. 
(a) $\phi=0$. (b) $\phi=-0.3\pi$.
}\label{fig5}
\end{figure}
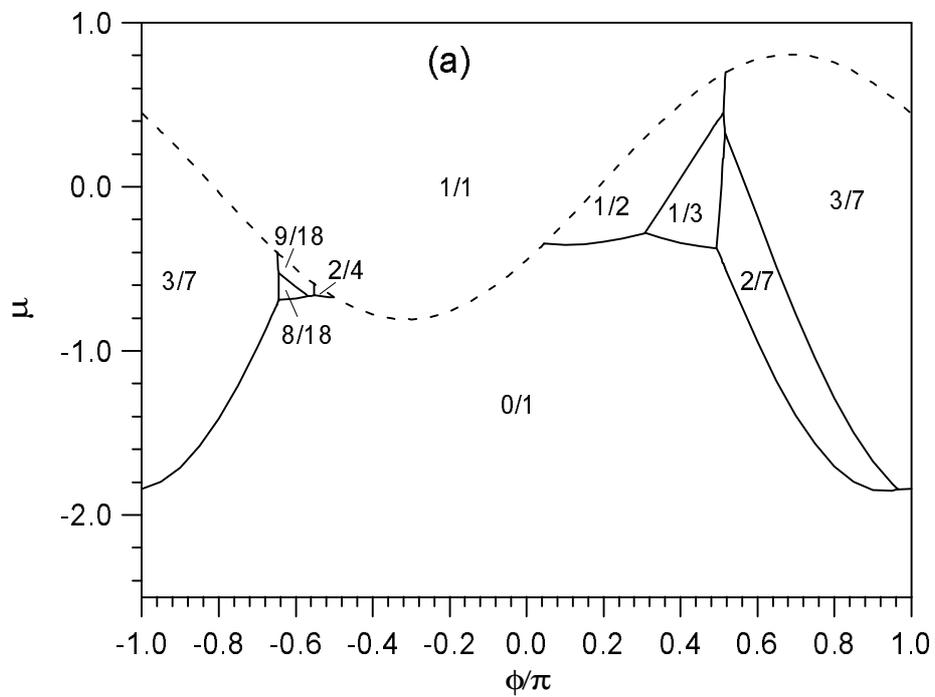
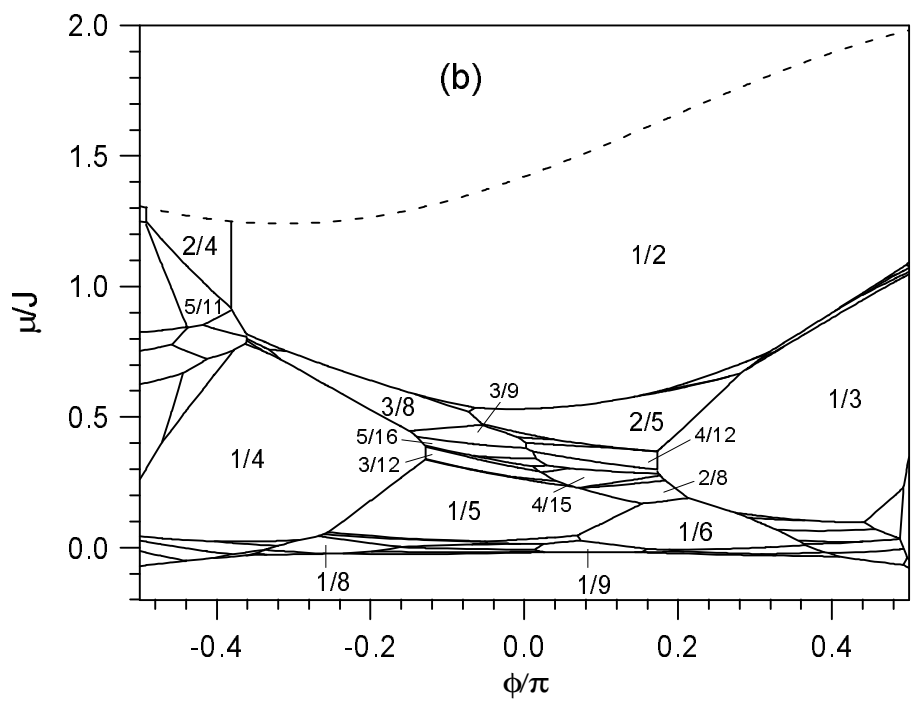
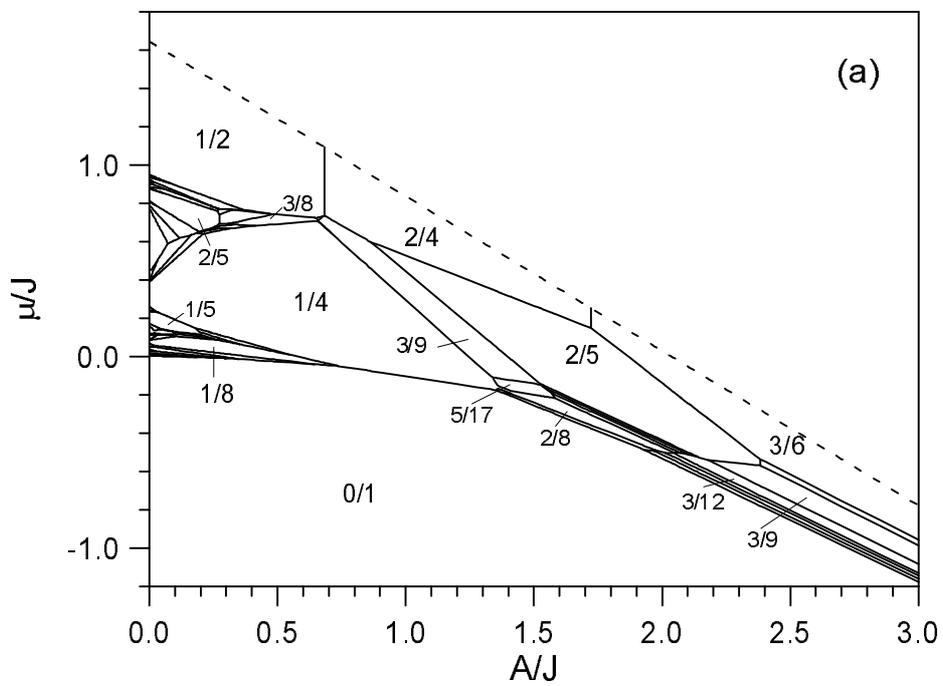
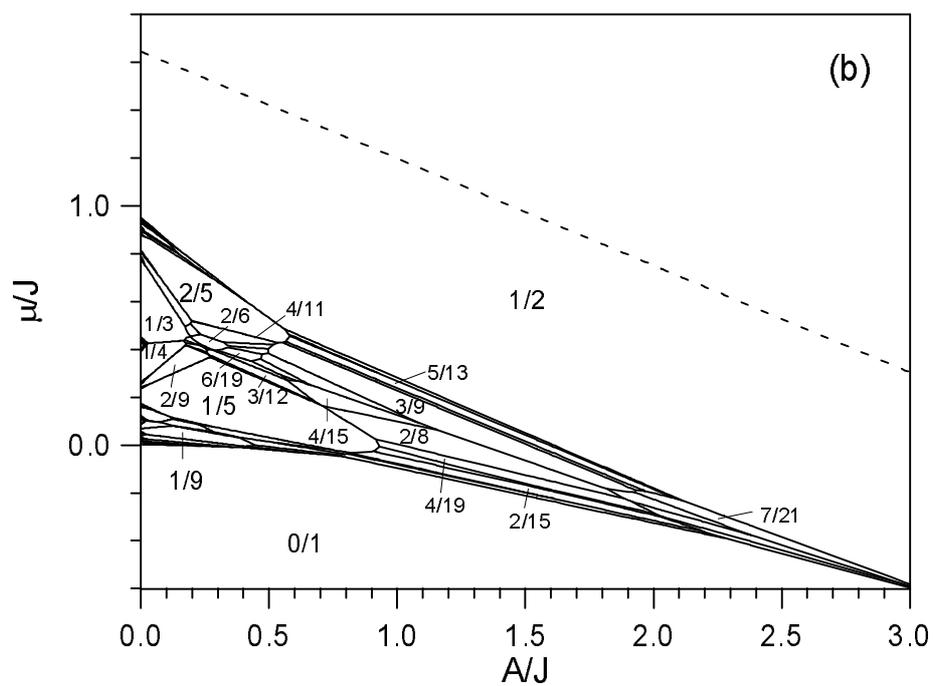
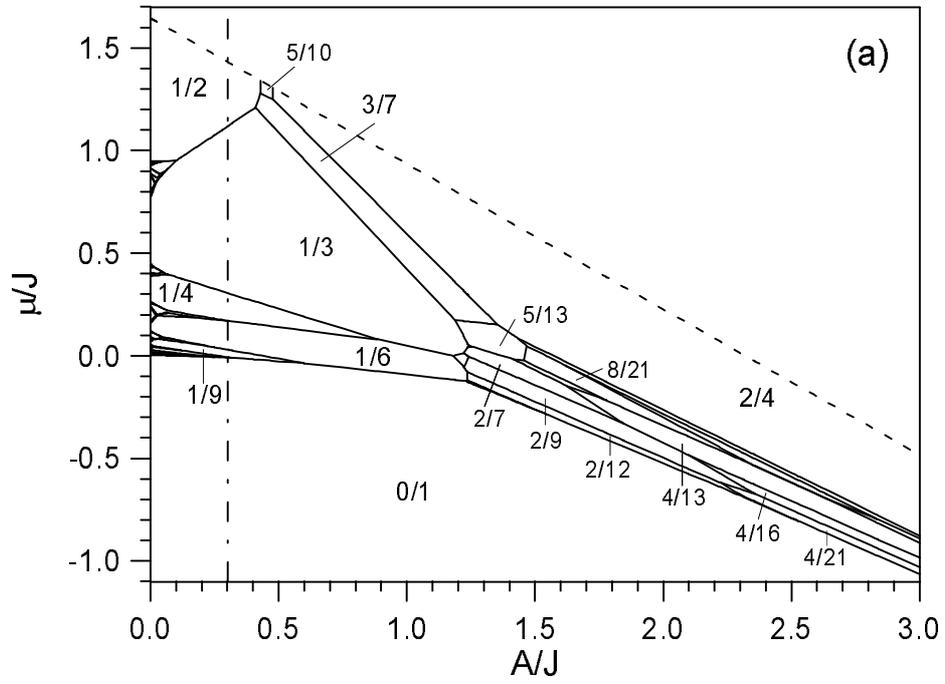
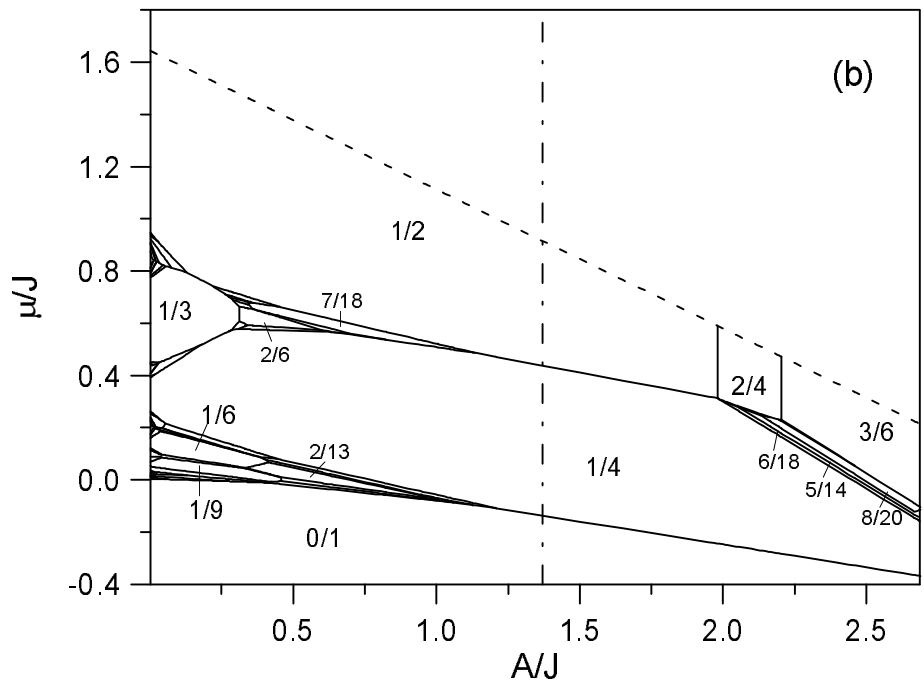
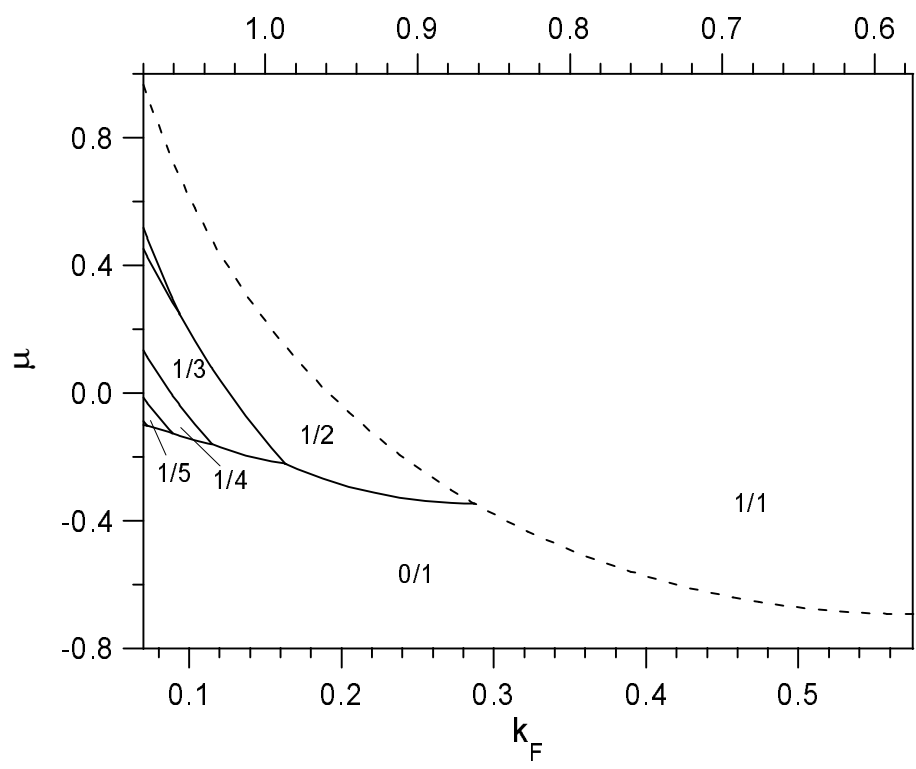
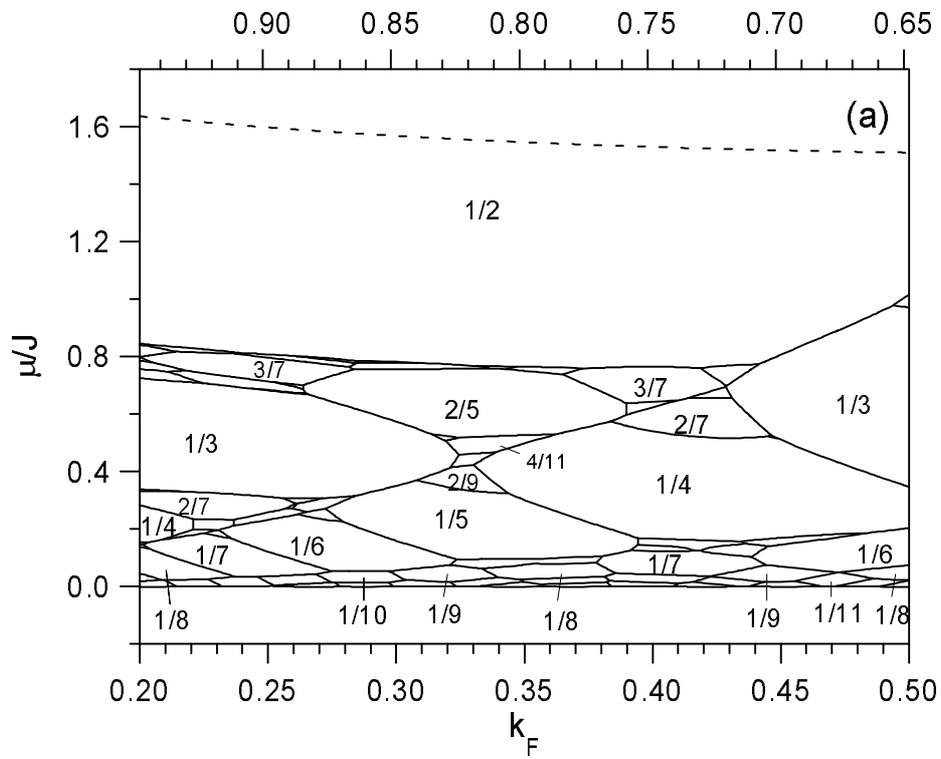
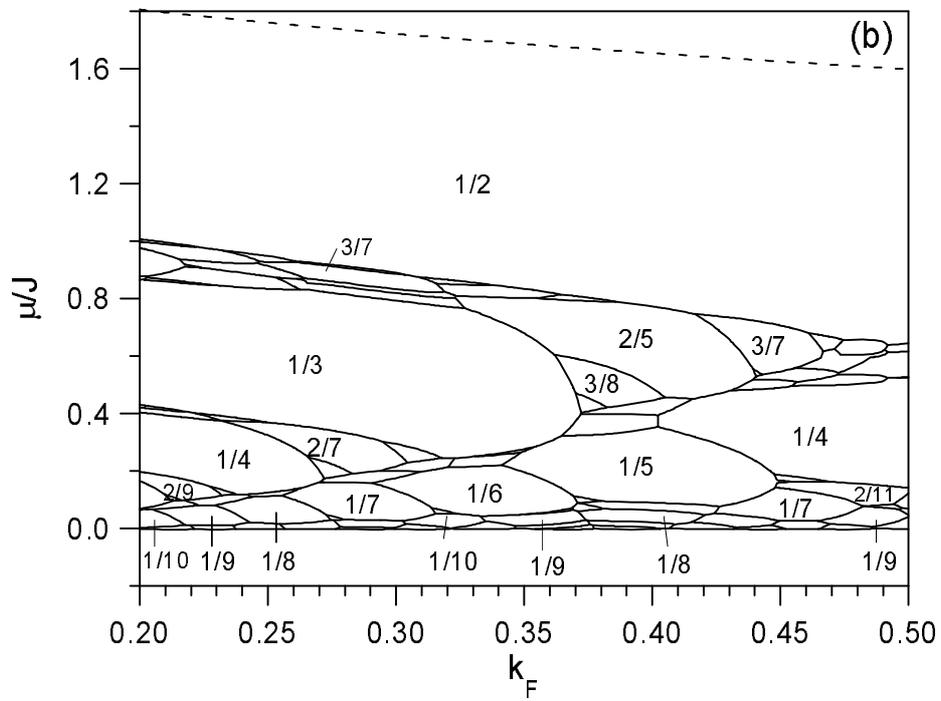
\end{document}